\documentclass[twocolumn,showpacs,aps,pra,amsfonts,amsmath,amssymb,superscriptaddress,floatfix]{revtex4}

\usepackage[dvips]{graphicx}

\newcommand{\tr}{{\rm \, Tr }\, }

\newcommand{\affA}{%
     Department of Applied Physics,
     The University of Tokyo, \\
     7-3-1 Hongo, Bunkyo-ku, Tokyo 113-8656, Japan}
\newcommand{\affB}{%
     Advanced Communications Technology Group,
     National Institute of Information and Communications Technology
     (NICT), \\
     4-2-1 Nukui-kitamachi, Koganei, Tokyo 184-8795, Japan}
\newcommand{\affC}{%
     Department of Electronics and Electrical Engineering, 
     Keio University, \\
     3-14-1 Hiyoshi, Kohoku-ku, Yokohama 223-8522, Japan}
\newcommand{\affD}{%
     CREST, Japan Science and Technology Agency, 
     1-9-9 Yaesu, Chuo-ku, Tokyo 103-0028, Japan}

\begin{document}
\title{7 dB quadrature squeezing at 860 nm with periodically-poled
KTiOPO$_4$}
%
\author{Shigenari Suzuki}
\affiliation{\affA}%
\affiliation{\affB}%
\affiliation{\affC}%
\author{Hidehiro Yonezawa}
\affiliation{\affA}%
\affiliation{\affD}%
\author{Fumihiko Kannari}
\affiliation{\affC}%
\author{Masahide Sasaki}
\affiliation{\affB}%
\affiliation{\affD}%
\author{Akira Furusawa}
\affiliation{\affA}%
\affiliation{\affD}%

\begin{abstract}
We observed $-7.2 \pm 0.2$ dB quadrature squeezing at 860 nm by using a
 sub-threshold continuous-wave pumped optical parametric oscillator with
 a periodically-poled KTiOPO$_4$ crystal as a nonlinear optical medium.
 The squeezing level was measured with the phase of homodyne detection
 locked at the quadrature.  
 The blue light induced infrared absorption was not observed in the
 experiment.  
\end{abstract}

\maketitle

Squeezed states of optical fields are important resources for photonic
quantum information technology particularly with continuous variables 
\cite{braunstein98, furusawa98, ban99, braunstein00-pra, ban00, li02,
ralph02, bowen03, mizuno05}.  
The performance of such protocols is limited directly by the squeezing
level \cite{braunstein98, takei05-prl}.  
For example, the fidelity in $n$ cascaded
quantum teleportation of coherent states scales as 
\begin{equation}
 F(n,r)=1/(1+n e^{-2r})
 \label{eq-F-r}
\end{equation}
with $r$ the squeezing degree \cite{furusawa98, braunstein00,
hammerer05}.  
The amount of information extracted by 
quantum dense coding must increase as 
\begin{equation}
 I(n_s,r)=\ln[1+n_s e^{2r}],
 \label{eq-I}
\end{equation}
where $n_s$ is the average photon number used for 
signal modulation \cite{braunstein00-pra, ralph02}.

The highest squeezing level observed so far 
under practical setting with the phase locked 
was $-6.0 \pm 0.3$ dB by Polzik {\it et al.} 
\cite{polzik92}. 
They employed a continuous-wave (CW) Ti:Sapphire laser at 852 nm and a
sub-threshold degenerate optical parametric oscillator (OPO) with a
KNbO$_3$ crystal as a nonlinear optical medium.  
Since then this scheme has been a sort of standard in squeezing
experiments at this wavelength range.

For KNbO$_3$, however, the pump (blue) light induced infrared absorption
(BLIIRA) has been known as the limiting factor for attaining higher
squeezing.  
A break-through was brought by Aoki {\it et al.} \cite{aoki05}, using 
periodically-poled KTiOPO$_4$ (PPKTP).  
Although existences of the pump light induced absorption in KTiOPO$_4$
(KTP) and PPKTP crystals had already been reported in pulsed light
experiments \cite{wang04}, BLIIRA was not observed at 946 nm in their CW
experiment \cite{aoki05}.  

In this letter, we report the higher level of squeezing, 
$-7.2 \pm 0.2$ dB, generated with PPKTP at 860 nm in CW experiment with
the phase locked.  
This squeezing level opens potential applications to new coding in
a single mode bosonic channel and sensing technologies
\cite{polzik92-prl, caves81, braunstein00-pra}.  
Particularly, it is expected to beat the Holevo capacity limit if the
state purity is improved by reducing the present anti-squeezing level
because the squeezing level itself already exceeds the theoretical
criterion of $-6.78$ dB \cite{braunstein00-pra}.  
The size and performance of photonic quantum circuit will also be
improved.  
For example, in the quantum teleportation of coherent states, either
five cascaded processes or a single process with a high fidelity of 0.84
could be performed in principle.  
Furthermore, the wavelength range corresponds to the Cs D$_2$ line 
(852 nm), 
and hence fascinating for applications for controlling Cs atoms with
non-classical light.

\begin{figure}[hbt]
 \includegraphics[width=85mm]{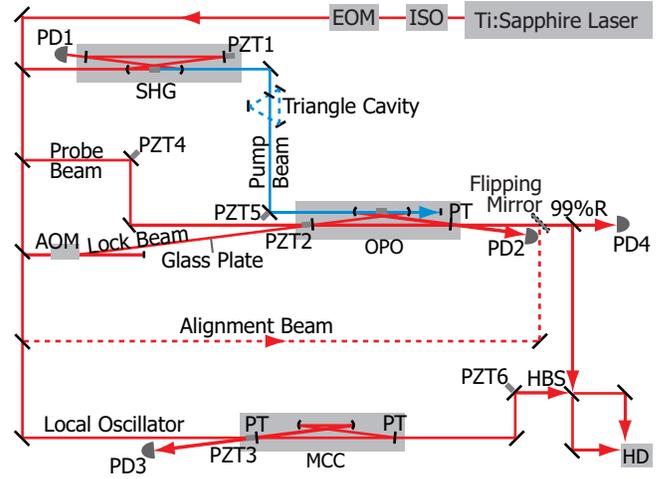}
 \caption{Experimental setup.  
 ISO: optical isolator, 
 EOM: electro-optic modulator, 
 AOM: acousto-optic modulator, 
 SHG: second harmonic generator (frequency doubler),
 OPO: sub-threshold degenerate optical parametric oscillator, 
 MCC: mode cleaning cavity, 
 HBS: 50:50 beam-splitter, 
 PTs: partial transmittance mirrors, 
 HD: balanced homodyne detector, 
 PDs: photo-detectors, 
 and PZTs: piezo-electric transducers.}
 \label{fig-setup}
\end{figure}
A schematic of our experimental apparatus is shown in
Fig. \ref{fig-setup}.  
A continuous-wave Ti:Sapphire laser (Coherent MBR-110) at 860 nm is
employed to accomplish this experiment.  
The beam from the Ti:Sapphire laser is phase-modulated at 15.3 MHz by
an electro-optic modulator after passing an optical isolator.  
The modulation is utilized to lock a cavity for frequency doubling and 
a mode cleaning cavity with conventional FM-sideband locking technique
\cite{drever83}.  

A part of the beam of around 900 mW is introduced into the frequency
doubler to generate second harmonic at 430 nm as a pump beam for an
OPO.  
The frequency doubler has a KNbO$_3$ crystal as a nonlinear optical
medium in the external cavity with a bow-tie-type ring configuration.  
An output power from the doubler at 430 nm is more than 400 mW.  

The OPO consists of a bow-tie-type ring cavity and a PPKTP crystal
(Raicol Crystals).  
The cavity has the folding angle of $7^\circ$, two spherical mirrors
(radius of curvature: 50 mm), and two plain mirrors.  
One of the plain mirrors is a partial transmittance mirror and works as
the output coupler while the others are high reflectance mirrors.  
The round-trip length of 500 mm, the distance between the two spherical
mirrors of 58 mm, and the PPKTP crystal (10 mm long) placed between the
spherical mirrors result in waist radii of 20 $\mu$m inside the crystal
and 200 $\mu$m outside the crystal.  
The cavity is mechanically stabilized by concatenating mirror mounts of
the mirrors with aluminum plates.  
The OPO easily oscillates with the pump power of 200 mW, while the
oscillation threshold $P_{th}=181$ mW is theoretically obtained from a
nonlinear coefficient of the crystal of $E_{NL}=0.023~ {\rm W}^{-1}$, an
intra-cavity loss $L=0.006$, and a transmittance of the output coupler of
$T=0.123$ with the formula $P_{th}=(T+L)^2/4E_{NL}$.  

The resonant frequency of the OPO cavity is locked using a ``lock
beam'' and a photo-detector (PD2) in Fig. \ref{fig-setup} via the
conventional FM-sideband locking technique \cite{drever83}.  
In order to avoid interference of the lock beam and a ``probe beam'' in
Fig. \ref{fig-setup}, the beams are in opposite circulations of the
cavity.  
Despite the effort, a fraction of the lock beam circulates backward
because of reflection from surfaces of the crystal.  
This problem is solved by changing the transverse mode and the frequency
of the lock beam \cite{polzik92}.  
The transverse mode is changed from TEM$_{00}$ to TEM$_{10}$ by
inserting a glass plate into a spatial half part of the beam.  
The frequency is shifted by about $-120$ MHz with an acousto-optic
modulator.   
Thus the probe beam in TEM$_{00}$ mode and the lock beam in TEM$_{10}$
mode resonate simultaneously with the OPO cavity.  

The generated squeezed light is combined with a local oscillator (LO) at
a 50:50 beam-splitter (HBS) and detected by a balanced homodyne detector
(HD) with Si photo-diodes (Hamamatsu S3590-06 with special
anti-reflection coating).  
The circuit noise level of the homodyne detector at 1 MHz is $-18.5$ dB
below the shot noise level with the LO of 3 mW.  
An output of the HD is measured for the sideband component at 1 MHz by a
spectrum analyzer (Agilent E4402B).  
The spectrum analyzer is set to the zero-span mode at 1 MHz with 30 kHz
resolution bandwidth and 300 Hz video bandwidth.  

\begin{figure}[tbh]
 \includegraphics[width=85mm]{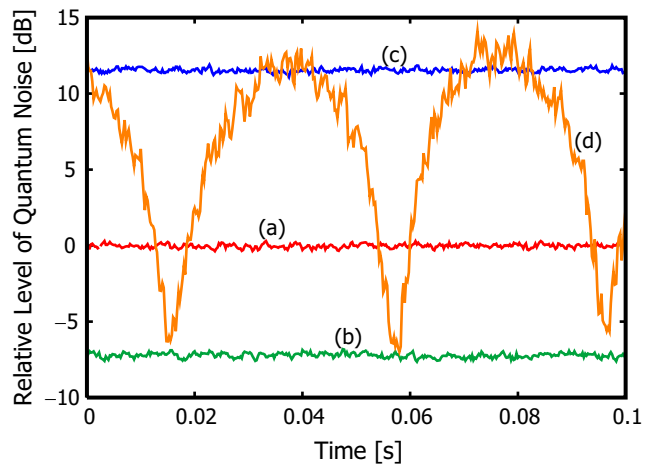}
 \caption{Power levels of quantum noise.  
 (a) Shot noise level.  
 (b) LO phase is locked at the squeezed quadrature.  
 (c) LO phase is locked at the anti-squeezed quadrature.  
 (d) LO phase is scanned.  
 These are normalized to make the shot noise level 0 dB.  
 All traces except for (d) are averaged 20 times.  
 }
 \label{fig-nl-time}
\end{figure}
A relative phase between the LO and the squeezed light is locked by
use of the probe beam.  
The probe is locked in phase so that it is minimized/maximized along
with the parametric gain of the OPO, and then works as a marker of
the squeezed/anti-squeezed quadrature of the squeezed light.  
As shown in Fig. \ref{fig-setup}, it is modulated in phase at 64 kHz by
a piezo-electric transducer (PZT4), amplified or deamplified in the OPO,
and then detected the fraction of 1 \% by a photo-detector (PD4).  
The phase modulation at 64 kHz is used for two controllers based on the
FM-sideband locking technique \cite{drever83}.  
An output signal of the PD4 is monitored by one of the controllers which
locks the probe via controlling a piezo-electric transducer (PZT5).  
Then, a relative phase between the probe and the LO is locked by the
other controller which monitors an interference signal of these beams
from the HD and controls another piezo-electric transducer (PZT6).  
A fluctuation of the relative phase between the LO and squeezed light is
estimated as $\tilde \theta = 3.9^\circ$, which is obtained via measuring
rms values of error signals from the control circuits.  

In order to improve the homodyne efficiency, the LO is spatially
filtered by the mode cleaning cavity which yields the same spatial mode
as the OPO output.  
The overall detection efficiency $\eta=\eta_P \eta_H$ after the OPO is
obtained from the propagation efficiency of the optical path of
$\eta_P=0.99$ and the homodyne efficiency $\eta_H=0.98$.  
The homodyne efficiency is dominated by the visibility between the LO
and the OPO output mode because the quantum efficiency of the Si
photo-diodes could be treated as unity at the wavelength.  

In addition, the alignment beam shown in Fig. \ref{fig-setup} is an
auxiliary beam which is reserved for use in alignment of constructing
the cavity, measuring the intra-cavity loss, and matching the spatial
mode of the pump beam with the OPO cavity.  
For the last application listed above, the alignment beam is converted
to the second harmonic with the OPO as a reference beam for the
alignment \cite{polzik92}.  
The reference beam propagates in the opposite direction to the pump beam
and represents the OPO cavity mode.  
By matching the spatial mode of the reference beam with that of the pump
beam, the pump beam is matched with the OPO cavity mode.  
In order to attain it, a triangle cavity in the path of the pump beam
is utilized \cite{polzik92}.

\begin{figure}[t]
 \includegraphics[width=85mm]{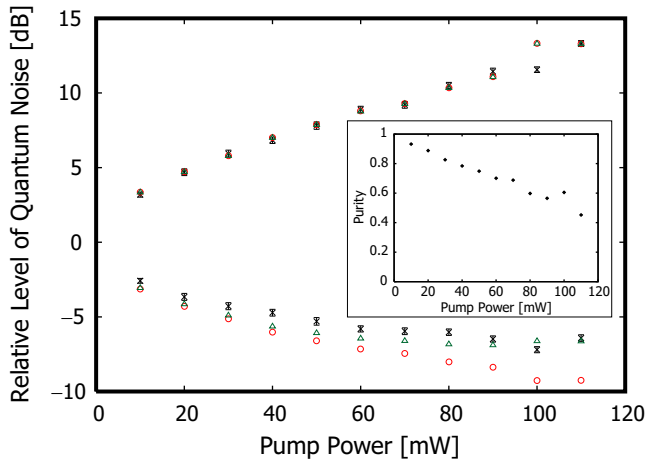}
 \caption{Squeezing and anti-squeezing levels at several powers of
 the  pump beam.  Plots with $\times$ indicate measured values while
 $\bigcirc$ and $\triangle$ indicate theoretical ones which are
 calculated from measured classical parametric gains.  The phase
 fluctuation of the LO is taken into account for the plots with
 $\triangle$ while it is not done for those with $\bigcirc$.  
 The inset shows purities calculated from the observed squeezing and
 anti-squeezing levels using the definition $\tr \{\hat \rho^2\}$ where
 $\hat \rho$ denotes the density operator of the observed state.} 
 \label{fig-sq-blue}
\end{figure}

Observed quantum noise levels of the OPO output are shown in
Fig. \ref{fig-nl-time}.  
They were observed when the pump power was 100 mW.  
We succeeded in locking the LO phase at the squeezed and anti-squeezed
quadratures and in obtaining  $-7.2 \pm 0.2$ dB squeezing and $+11.6 \pm
0.2$ dB anti-squeezing.  
The squeezing level of the generated state is estimated to be $-7.5$ dB
by taking into account the effect of the circuit noise.  

Note that the intra-cavity loss $L=0.006$ of the OPO stayed constant
independently of the pump beam power, i.e., BLIIRA was not observed.  
Considering the existence of the pump light induced infrared absorption
in experiments with pulsed lasers \cite{wang04}, we infer that the
absence of BLIIRA in our experiment is due to the CW operation.  

The pump power dependences of the squeezing and anti-squeezing levels
are shown in Fig. \ref{fig-sq-blue}.  
The observed squeezing level saturates while the pump power increases.  
This fact could be due to the fluctuation $\tilde \theta$ which had a
large effect on the observed squeezing level through mixing of the
highly anti-squeezed quadrature.  
Taking account of $\tilde \theta$, the theoretical squeezing level
$R^\prime_{-}$ and anti-squeezing level $R^\prime_{+}$ are calculated as
follows \cite{zhang03}: 
\begin{equation}
 R_{\pm}^\prime \approx
  R_{\pm} \cos^2 \tilde \theta +R_{\mp} \sin^2 \tilde \theta,
  \label{eq-rprime}
\end{equation}
where $R_{\pm}$ is modeled in Refs. \cite{polzik92,collett84}.  
Considering $\tilde \theta =3.9^\circ$, the theoretical results almost
agree with the experimental ones as shown in Fig. \ref{fig-sq-blue}.  
Assuming $\tilde \theta =0$, the squeezing level would be $-9.3$ dB with
100 mW pumping.

In conclusion, we achieved  $-7.2 \pm 0.2$ dB quadrature squeezing at
860 nm with the phase locked at the maximally squeezed quadrature.  
Moreover, BLIIRA was not observed in a PPKTP crystal for the case of the
CW pumped sub-threshold optical parametric oscillator.  
The resulting state is expected to be utilized to perform various kinds
of quantum information processing, to implement precise measurements,
and to investigate the photon-atom interactions with Cs atoms.

\begin{acknowledgments}
 SS is grateful to Nobuyuki Takei for his experimental support.  
 This work was partly supported by the MPHPT and the MEXT of Japan.  
\end{acknowledgments}

\end{document}